\documentclass[reprint,superscriptaddress,showpacs,amsmath,amssymb,twocolumn,prb]{revtex4-1}

\usepackage{graphicx}
\usepackage{dcolumn}
\usepackage{bm}
\usepackage[usenames, dvipsnames]{xcolor}
\usepackage{soul}
\usepackage{nicefrac}

\newcommand{\cmmnt}[1]{}

\begin{document}

\title{Step-like spectral distribution of photoelectrons at the percolation threshold\\ in heavily $p$-doped GaAs}

\author{S.~V.~Poltavtsev}
\email{sergei.poltavtcev@tu-dortmund.de}
\affiliation{Experimentelle Physik 2, Technische Universit\"at Dortmund, 44221 Dortmund, Germany}
\affiliation{Spin Optics Laboratory, St.~Petersburg State University, 198504 St.~Petersburg, Russia}
\author{R.~I.~Dzhioev}
\affiliation{Ioffe Institute, Russian Academy of Sciences, 194021 St.~Petersburg, Russia}
\author{V.~L.~Korenev}
\affiliation{Ioffe Institute, Russian Academy of Sciences, 194021 St.~Petersburg, Russia}
\author{I.~A.~Akimov}
\affiliation{Experimentelle Physik 2, Technische Universit\"at Dortmund, 44221 Dortmund, Germany}
\affiliation{Ioffe Institute, Russian Academy of Sciences, 194021 St.~Petersburg, Russia}
\author{D.~R.~Yakovlev}
\affiliation{Experimentelle Physik 2, Technische Universit\"at Dortmund, 44221 Dortmund, Germany}
\affiliation{Ioffe Institute, Russian Academy of Sciences, 194021 St.~Petersburg, Russia}
\author{M.~Bayer}
\affiliation{Experimentelle Physik 2, Technische Universit\"at Dortmund, 44221 Dortmund, Germany}
\affiliation{Ioffe Institute, Russian Academy of Sciences, 194021 St.~Petersburg, Russia}

\date{\today}

\begin{abstract}
We study the origin of the step-like shoulder on the high energy side of the low temperature photoluminescence spectrum of heavily $p$-doped GaAs. We show experimentally that it is controlled by the Fermi-Dirac distribution of the holes and by the energy distribution of the photoexcited electrons showing a sharp step-like dependence. This step is attributed to the percolation threshold in the conduction band separating localized from delocalized electron states. A comprehensive set of optical techniques based on spin orientation of electrons, namely the Hanle effect, time- and polarization-resolved photoluminescence, as well as transient pump-probe Faraday rotation are used for these studies. We identify two different electron ensembles with substantially different spin lifetimes of 20 and 280~ps, limited by the lifetime of the electrons. Their spin relaxation times are longer than 2~ns. The relative contribution of short- and long-lived photoexcited electrons to the emission spectrum changes abruptly at the high-energy photoluminescence step-like tail. For energies above the percolation threshold the electron states are empty due to fast energy relaxation, while for lower energies the relaxation is suppressed and the majority of photoelectrons populate these states.
\end{abstract}

\maketitle

\section*{Introduction}

In a high purity semiconductor, such as epitaxially grown GaAs, the light absorption is dominated by the narrow resonant line of excitons, which remain free even at liquid helium temperatures. When, however, the crystal is doped to a sufficiently high concentration of impurities, the absorption line broadens \cite{Madelung1996}. The reason is that the high impurity density disturbs the crystal translation symmetry introducing disorder. As a result, the carrier momentum conservation law is not strictly fulfilled and optical transitions indirect in $\bf k$-space become possible. This unavoidably leads to broadening of the sharp spectral features in photoluminescence (PL) and PL excitation (PLE) spectra.

There are several remarkable phenomena that take place in systems with strong disorder. For example, the conductivity increases rapidly in a step-like way for continuous increase of charge carriers, which is know as metal-insulator transition \cite{EfrosBook}. In particular, such step-like behavior takes place due to existence of the percolation threshold, which separates the localized and delocalized states in the energy spectrum \cite{Anderson1958, EfrosBook}. The percolation threshold exists in the empty band within the single particle Anderson model~\cite{Anderson1958}. Another prominent example for a step-like behavior in a disordered system is the quantum Hall effect, which is manifested by pronounced steps in the Hall conductivity when scanning the external magnetic  field \cite{QuantumHall}. These steps appear due to the presence of disorder. A further remarkable phenomenon, which occurs in the case of strong acceptor doping of direct band gap semiconductors, is manifested by the appearance of a sharp edge-like shoulder in the high energy flank of the PL spectrum at low temperatures \cite{CardonaPRB1980, MillerPRB1981}. So far, it has been associated with the formation of the Fermi-Dirac distribution of the holes in the valence/acceptor band and the sharpness of this edge has been believed to trace the hole temperature \cite{CardonaBook}.

However, the Fermi level of the holes alone is insufficient to explain the formation of the sharp low-temperature PL edge in a heavily $p$-doped crystal. An additional condition for the distribution of the photoexcited electrons in the conduction band is required. 

In this study we demonstrate that the energy distribution of these electrons in heavily $p$-doped GaAs is described by a sharp step-like function. The energy where the step occurs corresponds to the percolation threshold. For energies larger than the percolation threshold electron states are not occupied due to their fast energy relaxation. For lower energies the relaxation is suppressed and the majority of photoexcited electrons populates energy states just below the percolation threshold. Two important features of these observations deserve particular attention. First, the population of electrons is not described by a Fermi-Dirac distribution because the photoexcited electrons are not in equilibrium. Second, the statistics is far from being degenerate since the population is significantly smaller than unity, even for the electrons with energy below the percolation threshold. This is because the density of photoexcited electrons in typical PL experiments is about $10^{13}-10^{14}$~cm$^{-3}$, which is significantly smaller than the concentration of donors $10^{16}-10^{17}$~cm$^{-3}$, which are present due to compensation related to the growth.

We use a set of optical techniques based on optical orientation of electrons in semiconductors~\cite{OptOrientBook} in order to study the dynamics of photoexcited electrons in a highly $p$-doped GaAs crystal. Using the Hanle effect, i.e. the depolarization of PL in a transverse magnetic field, we evaluate the spin lifetime, which undergoes a drastic changes at photon energies close to the sharp PL edge. Using time-resolved PL and pump-probe Faraday rotation we probe the spin dynamics and evaluate the population and spin relaxation times for electrons at different energies. At low temperatures, strong changes of the electron lifetime for photon energies around the high energy PL edge demonstrate the presence of the percolation threshold for photoexcited electrons in the conduction band.

The paper is organized as follows. First, the energy level structure of the studied samples is considered to understand its PL spectrum. Then, the Hanle effect is studied in two ranges of magnetic field strengths. The kinetics of the photoexcited carriers and their spins is then studied by time-resolved PL and pump-probe Faraday rotation, followed by a discussion and conclusions.

\vspace{5mm}

\section*{Experimental results}

\subsection{Samples and photoluminescence}

The main studied sample (AH4227) is a 25 $\mu$m GaAs layer heavily $p$-doped by Ge with a concentration of $N_A=5\times10^{18}$~cm$^{-3}$ which was grown by metal-organic chemical vapor deposition (MOCVD). The GaAs layer is separated from the GaAs substrate by a 2 $\mu$m-thick Al$_{0.3}$Ga$_{0.7}$As layer. A 2~$\mu$m-thick cover layer of Al$_{0.3}$Ga$_{0.7}$As was deposited on top of the structure. Additionally, a similar structure (R94-1) with acceptor concentration of $N_A=3\times10^{18}$~cm$^{-3}$ was investigated.

\begin{figure}[t]
	\vspace{5mm}
	\includegraphics[width=\linewidth]{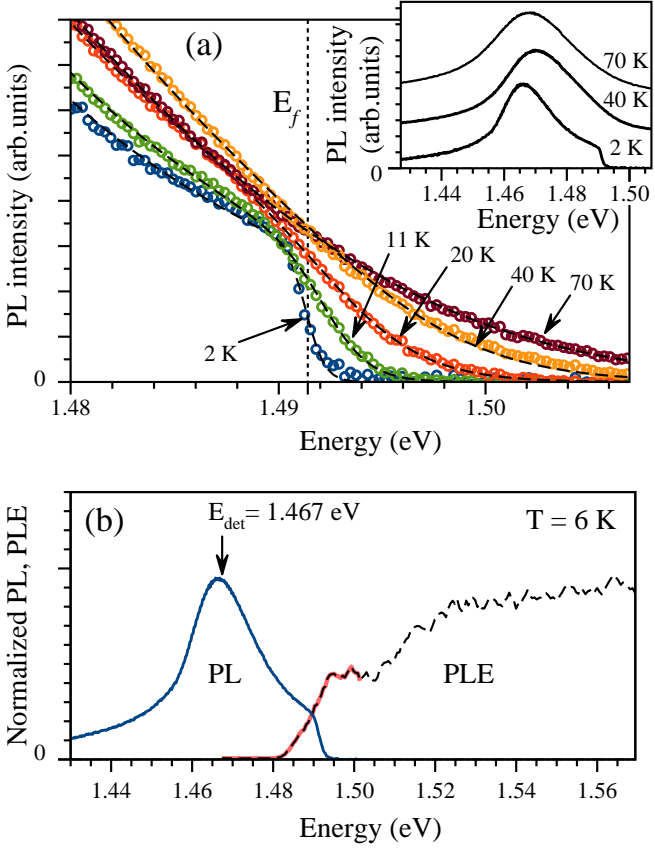}
	\caption{PL and PLE spectroscopy of $p$-doped GaAs with acceptor density $N_A= 5\times10^{18}$ cm$^{-3}$: (a) Smoothing of the step in the high energy side of the PL spectrum shoulder with rising sample temperature. Circles give experimental data; dashed lines are fits with Eq.~(\ref{PL_fit}). Inset shows complete PL spectra over an extended energy range for different temperatures, shifted vertically for clarity. (b) Comparison of normalized PL (solid line) and PLE (dashed line) spectra measured at $T=6$~K. PLE detection energy is 1.467~eV. The step-like feature in the PLE spectrum below 1.50~eV is highlighted.}
	\label{PL_PLE}
\end{figure}

A set of PL spectra measured at temperatures $T=$ 2--70 K under cw-laser excitation with the photon energy of 1.55~eV is shown in Fig.~\ref{PL_PLE}(a). The broad PL, even at low temperatures, features a characteristic sharp edge on the high energy side. This observation is in excellent correspondence with PL spectra of heavily $p$-doped GaAs reported previously by other authors \cite{CardonaPRB1980, MillerPRB1981, Hudait1998, Kucera2006}. The spectrum's shape does not change with increasing excitation density up to at least 30~W~cm$^{-2}$. The excitation density in most cw-experiments was within 10~W~cm$^{-2}$ and therefore below this value. Increasing the sample temperature results in a smoothing of the PL edge, as seen from Fig.~\ref{PL_PLE}(a). The shape of this edge can be fitted well with the following phenomenological function based on the Fermi-Dirac distribution: 

\begin{equation}
\label{PL_fit}
y(E)\sim e^{-E/E_W} \large/ \large(e^{(E-E_f)/k_BT_c}+1\large). 
\end{equation}

\noindent Here, $E_f\approx1.4914$ eV is an energy gap, which is weakly affected by temperature up to 30~K; $k_B$ is Boltzmann constant, and $E_W=0.016\pm0.002$ eV ($T=2-30$~K) is a fit parameter describing the high energy PL profile below the sharp edge which shows an exponential decay. The fitted hole temperature $T_c$ is in fairly good correspondence with the lattice temperature. This approach was used previously to explain the spectral profile of high energy PL edge in heavily doped $p$-type materials \cite{CardonaBook}.

To identify the absorption edge, the PLE spectrum was measured at $T=6$ K for a detection energy $E_{det}=1.467$~eV. The spectrum in Fig.~\ref{PL_PLE}(b) demonstrates also a step at the high energy edge of the PL spectrum. We conclude that below this photon energy the acceptor states are occupied with holes. Taking into account that, at a given acceptor concentration ($N_A=5\times10^{18}$ cm$^{-3}$), the metal-insulator transition takes place in GaAs \cite{Ferreira2004}, we confirm that it is the hole energy distribution that contributes to the characteristic edge-like tail in the PL spectrum. However, we highlight that there is a drastic difference in the width of the step in the PL and PLE spectra. In PL, the width at $2-10$~K is in the range of $0.2-0.4$~meV, while in PLE at 6~K the width is enlarged to 1~meV. If only the hole distribution would determine the spectrum shape, we would expect the same step widths from PL and PLE spectroscopy. Therefore an additional factor has to be taken into account for explanation of the sharp edge in PL spectrum.

\subsection{Hanle measurements}

In order to identify the spin lifetime $T_S$ of the photoexcited electrons in the vicinity of the step in the PL spectrum we record the PL depolarization by an externally applied transverse magnetic field, i.e. we measure the Hanle effect at various detection energies around the step. Note that the resident holes are not spin-polarized and do not contribute to the observed optical spin orientation.

Application of the magnetic field in Voigt geometry causes a high energy shift of the edge-like PL tail. For that reason we work in two ranges of magnetic field strength corresponding in one case to a relatively weak, negligible energy shift and in the other case to a non-negligible shift.

\subsection*{Voigt magnetic fields $< 0.5$~T}

In the first case, the standard Hanle method allowing for measurement of the spin lifetime $T_S = (1/\tau_S + 1/\tau)^{-1}$ through the width of the PL depolarization curve as function of the external magnetic field was used. Here, $\tau_S$ is the spin relaxation time and $\tau$ is the electron lifetime both of which limit the spin lifetime. The cw-excitation by light with photon energy of 1.55~eV and pump density of 6~W~cm$^{-2}$ was periodically modulated between the two counter-circular polarizations using a photoelastic modulator (PEM) at 50~kHz frequency. The degree of circular polarization (DCP) of the PL was measured by detecting the PL in $\sigma^+$ polarization using a spectrometer with resolution of $\sim1$~nm, equipped with a photomultiplier. The two-channel photon counter synchronized with the PEM was used to measure the PL intensities $I_{\sigma^+}$ and $I_{\sigma^-}$ emitted during the excitation periods with polarization according to the subscript. The DCP is calculated as

\begin{equation}
\rho_c=\frac{I_{\sigma^+} - I_{\sigma^-}}{I_{\sigma^+} + I_{\sigma^-}},
\label{DCPPL}
\end{equation}

\begin{figure}[t]
\vspace{5mm}
\includegraphics[width=\linewidth]{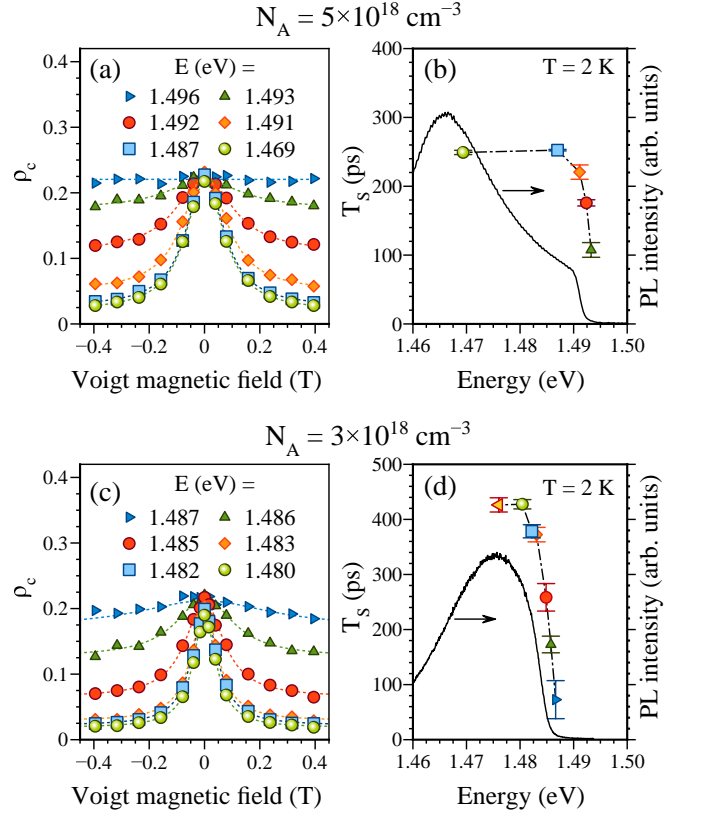}
\caption{PL depolarization measurements in Voigt magnetic fields $B<0.5$~T at $T=2$~K. Data are shown for the sample with acceptor density $N_A=5\times10^{18}$~cm$^{-3}$: (a) Set of PL depolarization curves measured as function of magnetic field strength for different detection energies (symbols). Fits based on Eq.~(\ref{Hanle_fit}) are shown with dashed lines. (b) Spectral dependence of the spin lifetime (symbols with error bars). The data points are connected with dash-dotted line for clarity. For comparison, the PL spectrum at $T=2$~K is shown by the solid line. The corresponding set of data for the sample with acceptor density $N_A=3\times10^{18}$~cm$^{-3}$ is displayed in panels (c)--(d).}
\label{Hanle_weak}
\end{figure}

\noindent and was measured as a function of the magnetic field strength up to 0.4~T for a set of detection energies at $T=2$~K. Each of these data sets, shown in Fig.~\ref{Hanle_weak}(a) by the symbols, was fitted by a zero-field centered Lorentzian  with a vertical offset, which corresponds to an additional contribution with very broad linewidth and accordingly short $T_S$. The PL depolarization profiles show drastic changes when moving the detection energy across the step-like feature in the PL spectrum: The larger the energy is, the broader is the profile and the higher is the offset. At the very tail of the PL line, where the PL signal becomes small ($E>1.495$~eV), the measured data are at a practically constant level. 
On the other hand, within the PL line ($E=1.469$~eV) the PL depolarization curve is quite narrow, which is the case for all energies within the broad PL spectrum.

From the fits to the data shown in Fig.~\ref{Hanle_weak}(a) by the lines, the spin lifetime $T_S$ of the relatively slow component corresponding to the narrow Hanle peak can be extracted using the equation

\begin{equation}
T_S =\frac{\hbar}{|g| \mu_B B_{1/2}}.
\label{Hanle_fit}
\end{equation}

\noindent Here, $\mu_B$ is the Bohr magneton, $\hbar$ is the Planck constant, $B_{1/2}$ is the Lorentzian half-width and $|g|=0.53$ is the electron $g$-factor measured by time-resolved PL, as will be shown below. The results of this analysis are shown in Fig.~\ref{Hanle_weak}(b), showing the spin lifetime as function of detection energy together with the PL spectrum at $T=2$ K. Within the PL line, the spin lifetime is $250\pm20$~ps, but it becomes substantially shorter at $E>1.49$~eV.

The same type of measurements was carried out also on another sample with the acceptor concentration of $N_A=3\times10^{18}$~cm$^{-3}$. The corresponding PL depolarization curves and extracted spin lifetimes $T_S$ together with the PL spectrum at $T=2$~K are displayed in Figs.~\ref{Hanle_weak}(c)--\ref{Hanle_weak}(d). This sample demonstrates a similarly sharp edge at the high energy side of the PL spectrum and shows a qualitatively similar broadening of the magnetic PL depolarization curves above this edge. The spin lifetimes drop from $T_S=$ 420~ps to 70~ps at the edge. We concentrate in the following on the sample with the acceptor concentration of $N_A=5\times10^{18}$~cm$^{-3}$.

We note that previous experiments on similar structures at temperatures of 77 and 300~K revealed only a small difference in the halfwidth of the magnetic depolarization curves (Hanle effect) \cite{OptOrientBook} between the short-wavelength and long-wavelength spectral ranges. This was caused by the influence of the diffusive electron drain from the excited surface into the depth of the sample \cite{Pierce1975, Dzhioev1993}.

\begin{figure}[t]
	\vspace{5mm}
	\includegraphics[width=\linewidth]{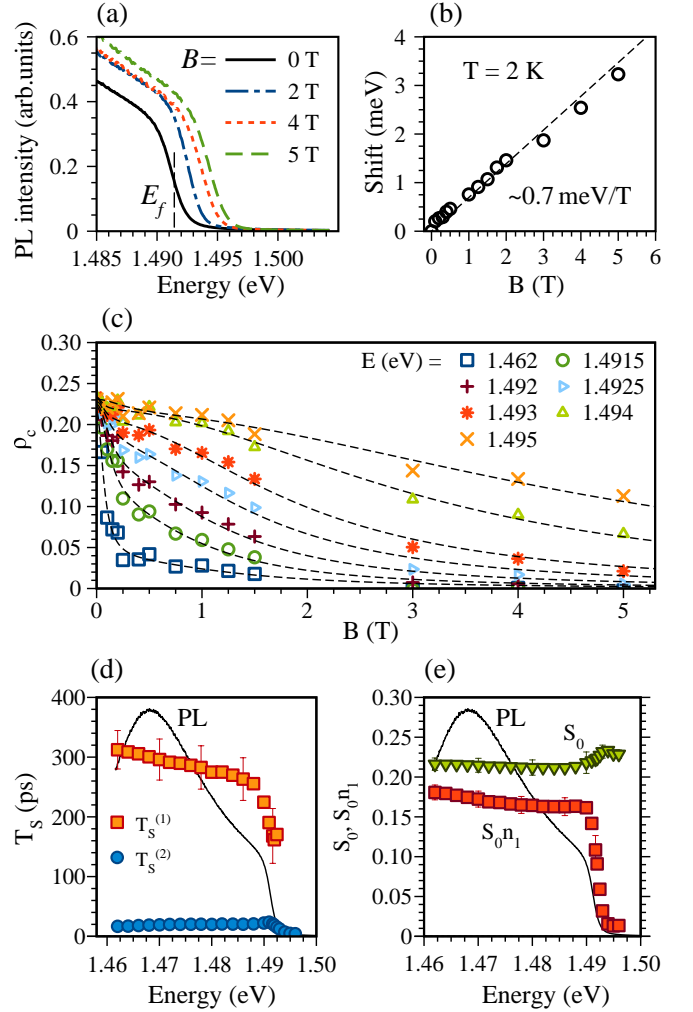}
	\caption{Hanle effect in strong magnetic fields: (a) Shift of the high-energy PL edge with increasing magnetic field strength. (b) High energy shift of the PL edge as function of magnetic field (circles), fitted with a $B$-linear dependence (dashed line). (c) Magnetic PL depolarization curves obtained by the technique described in the text (symbols) fitted by two Lorentzians (dashed lines); see Eq.~(\ref{Hanle_double_fit}). (d) Spin lifetimes of the two electron contributions: the localized electrons with $T_S^{(1)}\approx280$ ps (squares) and the free electrons with $T_S^{(2)}\approx20$ ps (circles). (e) Spectral dependence of the DCP, $S_0$ (triangles), and of the contribution of the localized electrons, $S_0n_1$, to $S_0$ (squares), at $B=0$. In panels (d) and (e) the error bars indicate the fitting error, and the PL spectrum at $T=2$~K is shown by the solid line.}
	\label{Hanle_strong}
\end{figure}

\subsection*{Voigt magnetic fields $> 0.5$~T}

Stronger magnetic fields are required to measure the short spin lifetimes, but they cause a significant high energy shift of the PL line. This can be seen from Figs.~\ref{Hanle_strong}(a)--\ref{Hanle_strong}(b), where in (a) the high energy edge of the PL spectrum measured at $T=2$~K is shown for different magnetic fields, while (b) shows the high energy shift of the edge as function of magnetic field. The high energy shift is proportional to the field strength with the slope given by $\sim0.7$~meV/T. This slope is somewhat smaller than the one expected for the energy shift of the electron level $\hbar\omega^e_c/2B\approx0.86$~meV/T, where $\omega^e_c$ is the electron cyclotron frequency. Due to this high energy shift, the measurement of the PL depolarization curves around the PL edge is nontrivial using the standard Hanle technique. Instead, the PL depolarization by the magnetic field was studied through measuring PL spectra at $T=2$~K. First, a set of PL spectra was recorded for different magnetic field strengths for $\sigma^+$ polarized excitation. The PL spectra were detected both for $\sigma^+$ and $\sigma^-$ polarization and then DCP spectra were calculated using Eq.~(\ref{DCPPL}). Then, the high energy shift of these spectra was compensated to have the energy gap $E_f$ at the same spectral position regardless of the magnetic field strength. Finally, cuts through this set of DCP spectra were taken for a number of fixed energies $E$, resulting in a set of data similar to the PL depolarization curves. A fraction of these data sets is displayed in Fig.~\ref{Hanle_strong}(c). They were fitted with the following expression containing the contributions from two types of electrons at a certain energy:

\begin{equation}
\rho_c=S_0\left[\frac{n_1}{1+\left(g\mu_BT_S^{(1)}B/\hbar\right)^2}+\frac{1-n_1}{1+\left(g\mu_BT_S^{(2)}B/\hbar\right)^2}\right].
\label{Hanle_double_fit}
\end{equation}

\noindent Here, $T_S^{(i)}$ are spin lifetimes of the slow ($i=1$) and fast ($i=2$) decaying components, $n_1$ and $1-n_1$ are corresponding weights of these components, and $S_0$ is the total DCP at $B=0$. The fitted values of $T_S^{(i)}$ and $n_1$are shown as functions of energy in Fig.~\ref{Hanle_strong}(d) and Fig.~\ref{Hanle_strong}(e), respectively. There is agreement between the spin lifetimes for the slow component measured with the two different techniques. 

The main result here is the identification of two types of electrons with substantially different spin lifetimes: $T_S^{(1)}\approx280$~ps and $T_S^{(2)}\approx20$~ps. Both types of electrons are observed across the whole PL spectral range. However, the PL line is mostly contributed by the electrons with the longer spin lifetime. At the high-energy PL edge, the weights of the two contributions de facto are exchanged, resulting in a dramatic shortening of the observed spin lifetime. Moreover, the spin lifetimes of both types of electrons experience a shortening at this edge. The sum of both contributions in the absence of magnetic field, $S_0$, also shown in Fig.~\ref{Hanle_strong}(e), exhibits no significant changes with energy variation and does not drop below 0.21. This value is close to the maximum achievable DCP value of 0.25 in GaAs, which, in turn, indicates that the spin relaxation time $\tau_S$ is much longer than the electron lifetime $\tau$. Based on the expression $S_0=0.25 / \left( 1 + \tau/\tau_S \right)$, the following estimation can be done: $\tau_S>5\tau$. Taking into account that the spin lifetime $1/T_S = 1/\tau + 1/\tau_S$, we conclude that $T_S$ is likely limited by the electron lifetime $\tau$.

\subsection{Time-resolved photoluminescence}

\begin{figure}[t]
	\vspace{5mm}
	\includegraphics[width=\linewidth]{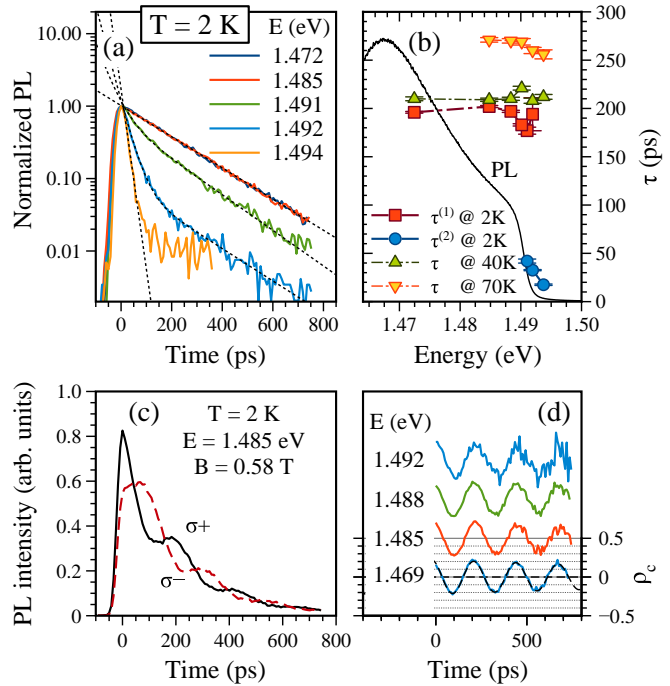}
	\caption{Time-resolved PL measurements: (a) Normalized PL kinetics measured with a streak-camera for different detection energies at $T= 2$ K (solid lines), fitted using mono- or bi-exponential decay forms (dashed lines). (b) Spectral dependence of the lifetime $\tau$ across the PL line, extracted from the fits to the PL kinetics measured at $T=2$, 40 and 70~K; the PL spectrum for $T= 2$~K is shown by the solid line. (c) PL transients measured at $E=1.485$~eV detection energy for $T=2$~K in an applied Voigt magnetic field of $B=0.58$~T. The excitation is $\sigma^+$-polarized, the detection is done in $\sigma^+$ and $\sigma^-$ polarization, as shown by the solid and dashed lines, respectively. (d) Spin dynamics $\rho_c$ detected for various detection energies (shifted vertically for clarity). The fit to the oscillations for $E=1.469$~eV detection energy is shown by the dashed line.}
	\label{TRPL}
\end{figure}

Time-resolved PL (TRPL) measurements using a streak camera were carried out in order to extract the electron lifetime $\tau$. Here, photoexcitation was done using picosecond laser pulses from a Ti:sapphire oscillator tuned to an energy of about 1.6~eV. The PL kinetics were detected spectrally resolved with a temporal resolution of about 20~ps and the data were averaged across spectral window of 0.5 nm. The PL kinetics measured at $T=2$ K for different energies $E$ in absence of the external magnetic field are shown in Fig.~\ref{TRPL}(a). They reveal drastic changes in the decay when crossing the high-energy PL edge. At the very edge ($E\gtrsim1.49$~eV) the PL decay contains two components with different decay times of about 20 and 200~ps, while within the PL line ($E<1.49$~eV) the fast component is absent. The PL kinetics were also measured at $T=40$~K and 70~K and exhibit purely mono-exponential decays. The data were accordingly fitted with either a single or bi-exponential function, the fitting results are displayed in Fig.~\ref{TRPL}(b). The shortest PL lifetime of $\tau^{(2)}\approx20$ ps, measured at $T=2$ K in the high-energy PL tail, is limited by the streak camera resolution. In the region $E<1.49$ eV, the PL lifetime is independent of the detection energy. The fast component vanishes with rising temperature and the long PL lifetime increases from about $\tau^{(1)}=200$ ps at $T=2$~K to about 270~ps at 70~K. 

Detection of the DCP $\rho_c(t)$ with temporal resolution makes it possible to measure the spin dynamics of the photoexcited carriers \cite{Heberle1994, AkimovPRB2009}. $\rho_c(t)$ was measured in an external magnetic field of $B=0.58$~T in Voigt geometry. Excitation was done with $\sigma^+$ polarized pulses and the PL kinetics were measured both in $\sigma^+$ and $\sigma^-$ polarizations ($I_{\sigma^\pm}$). These kinetics are shown in Fig.~\ref{TRPL}(c) for $E=1.485$~eV detection energy. The resulting DCP transients calculated using Eq.(\ref{DCPPL}) are plotted in Fig.~\ref{TRPL}(d) for a set of detection energies. All traces manifest oscillations, which can be fitted well with $\rho_c=A\cos(\Omega_Lt+\phi_0)\exp(-t/T_2^*)$. Here, $\Omega_L=g\mu_BB/\hbar$ is the Larmor frequency of electron spin precession, $A$ is the initial DCP, $g$ is the carrier $g$-factor, $T_2^*$ is the spin dephasing time, and $\phi_0$ is a phase offset. From these fittings we obtain $A=0.23\pm0.01$, which is close to the maximal possible DCP value $\rho_c$($B$=0$)=0.25$. The value of the $g$-factor $0.53\pm0.01$ is close to that of the electron, $|g|=0.44$, measured in pure GaAs \cite{WeisbuchPRB1977, OestreichPRL1995}. Based on this $g$-factor value, we confirm that the PL signal arises from the recombination of photoexcited electrons with  equilibrium holes. The obtained spin dephasing time of $T_2^*=2.0\pm0.7$~ns is limited either by the spin relaxation time $\tau_S$ or by a substantial electron $g$-factor dispersion. To identify the origin for this spin dephasing, the electron $g$-factor dispersion $\Delta g$ has to be measured.

\begin{figure}[t]
	\vspace{5mm}
	\includegraphics[width=\linewidth]{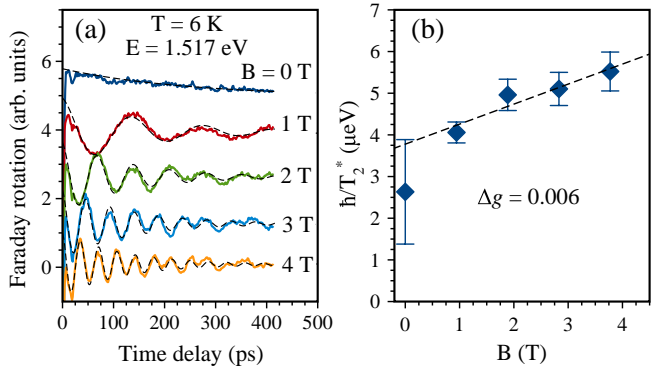}
	\caption{Time-resolved Faraday rotation at $T=6$~K for $E=1.517$~eV detection energy: (a) Faraday rotation signals at different Voigt magnetic fields (solid lines) with corresponding fits (dashed lines). (b) Extracted values of the spin dephasing rate $\hbar/T_2^*$ (symbols) as function of the magnetic field, shown with a $B$-linear fit function, from which $\Delta g=0.006$ is obtained.}
	\label{Faraday}
\end{figure}

\subsection{Time-resolved Faraday rotation}

Time-resolved pump-probe Faraday rotation was applied for accurate measuring the $\Delta g$ dispersion value. For this purpose, Voigt magnetic fields in the range  $0 - 4$~T were applied to the sample at $T=6$~K. Optical pumping and probing were done with two degenerate picosecond pulse trains, generated by a Ti:sapphire laser. The delay between the pump and probe pulses was varied by means of an optical delay line. The circular polarization of the pump pulse was modulated between $\sigma^+$ and $\sigma^-$ by a PEM at 50~kHz frequency. The rotation angle of the linear polarization of the probe pulse was measured in transmission geometry at this modulation frequency, using a detection scheme with a polarized beam-splitter and a balanced photodetector. 

The transients of Faraday rotation measured at $E=1.517$~eV excitation energy for different magnetic fields are shown in Fig.~\ref{Faraday}(a). In the absence of a magnetic field we observe a decay with a time constant of $T_0\approx250$~ps. This value agrees with the PL decay time measured on the high energy flank of the PL line ($\sim200$~ps at $E=1.492$~eV) and, thus, corresponds to the recombination time of the photoexcited electrons. 

The other transients for $B>0$ are fitted with exponentially decaying harmonic functions. This gives us the magnetic field-dependent time decay constant $T_2^*$, shown in Fig.~\ref{Faraday}(b) as the spin dephasing rate $\hbar/T_2^*$. At $B>0$ it can be approximated with a linear dependence $\hbar/T_2^*=\hbar/T_0 + \sqrt{2}\mu_B\Delta gB$, from which the value of the $g$-factor dispersion $\Delta g = 0.006$ is obtained. Using this value we estimate a spin dephasing time of about 2~ns at $B=0.58$~T. This agrees with the $T_2^*$ (also $\approx2$~ns) obtained from the TRPL experiment. Thus, we conclude that $\tau_S$ is at least several nanoseconds in the studied sample. We confirm also that the fitted electron $g$-factor $g\approx0.50$ value is in agreement with the TRPL measurements. Experimentally we checked that the $g$-factor does not depend on the temperature in the range $6-25$~K.

\section*{Discussion}

The most striking result here is the step-like spectral dependence of the halfwidth $B_{1/2}$ of the magnetic PL depolarization curve and its correlation with the sharp edge of the PL spectrum with a width $\Gamma<1$~meV (at $T=2$~K) in a strongly disordered semiconductor. Indeed, at a typical compensation degree of 0.1, the density of ionized donors and acceptors reaches the value $N_{D+}=N_{A-}\sim5\times10^{17}$~cm$^{-3}$. This means that the characteristic spatial potential fluctuations of the charged impurities are $\mathcal{E}_{fl}\sim (N_{D+})^{1/3}e^2/\kappa\approx5$~meV, where $e$ is the electron charge and $\kappa\approx13$ is the static dielectric constant in GaAs. Thereby, the bottom of the conduction band (as well as the top of the valence band) is broadened much stronger compared to $\Gamma$. 

\begin{figure}[h]
	\vspace{5mm}
	\includegraphics[width=\linewidth]{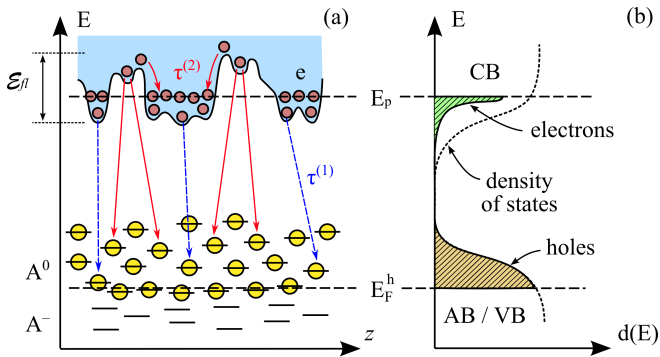}
	\caption{Energy states and optical transitions in a heavily $p$-doped GaAs crystal. (a) Possible optical transitions in the crystal. The potential landscape is shown with the curved line about the $E_p$ level. Notations: $e$ are electrons (small circles); $A_0$ are the neutral acceptors; $A^-$ are the ionized acceptors. Red solid and blue dashed arrows indicate the fast energy relaxation of the delocalized electrons ($\tau^{(2)}\approx20$~ps) and the slow recombination of the localized electrons ($\tau^{(1)}\approx280$~ps) with the equilibrium holes, respectively. $\mathcal{E}_{fl}$ is the characteristic energy scale of the potential fluctuations in the conduction band. (b) Density of states $d(E)$ (dotted line) and occupied states (shaded areas). $E_F^h$ is the Fermi level for the holes in the acceptor or valence band (AB/VB); $E_p$ is the percolation threshold for the electrons in the conduction band (CB).}
	\label{Scheme}
\end{figure}

Olego and Cardona suggested that the high energy edge in the PL spectrum is determined by the Fermi level delimiting the occupied hole states from the unoccupied ones \cite{CardonaPRB1980}. However, the momentum $\bf k$ in such a disordered system is not a good quantum number anymore and the optical transitions are not purely vertical in $\bf k$-space. Thus the position of the step is not well defined and is expected to be given by the inhomogeneous broadening of the conduction band edge. Also, the PLE spectrum [Fig.~\ref{PL_PLE}(b)] displays a smooth spectral behavior and does not demonstrate a sharp step-like edge as observed in the PL spectrum, despite the Fermi level for the holes. Note that at the studied shallow acceptor concentration of $N_{A^0}\sim5\times10^{18}$~cm$^{-3}$ the holes in GaAs are in the metallic phase of the metal-dielectric transition according to the Mott criterion \cite{MottBook}: $N_A^{1/3}a_H>0.25$, where $a_H\approx3$~nm is the hole Bohr radius on an acceptor \cite{OptOrientBook}. Thereby, assuming only the Fermi distribution step for carriers of one type is insufficient to explain the step-like edge in the PL spectrum, as has been believed so far. Consequently, explaining both the PL spectrum and the spectral behavior of the Hanle curve width $B_{1/2}$ requires an additional condition to be fulfilled, since the optical recombination of electrons that are continuously distributed in the conduction band would lead to a smooth spectral behavior of both the PL and $B_{1/2}$.

We suggest that this condition is the presence of a percolation threshold in the empty conduction band \cite{Anderson1958, EfrosBook}, i.e. of the energy level $E_p$, separating the localized electron from the extended electron states. Electron states with energies $E<E_p$ are spatially localized, while those with $E>E_p$ are delocalized over the crystal. The photoexcited electrons, which undergo rapid energy relaxation by emitting acoustic phonons, reach the $E_p$ energy level and quickly get trapped at this level, forming a step-like spectral distribution $n(E-E_p)$, as illustrated in Fig.~\ref{Scheme}(b). The energy relaxation by phonon emission occurs in small energy portions of $\hbar\omega_q=\hbar sq \leq k_BT\approx0.2$~meV, since the characteristic electron momentum transferred to the phonons is $q\sim1/L\sim10^6$~cm$^{-1}$. Here, $s\approx3\times10^5$~cm~s$^{-1}$ is the speed of sound in GaAs and $L\sim(N_{D+})^{-1/3}\sim10$~nm is the characteristic length scale of the potential fluctuations of the impurities. The characteristic energy relaxation time corresponds to the short time of $T^{(2)}_S=20$~ps, measured using the Hanle effect and TRPL. 

The energy relaxation rate slows down drastically at $E\leq E_p$, since the changes in electron energy due to the phonon emission are connected with changes of the electron position. In this case, the lifetime of electrons below the percolation threshold becomes significantly longer and is limited now by their radiative recombination with the holes at the acceptors, leading to an extension to $T^{(1)}_S=280$~ps. As a result, the recombination of the electrons below the $E_p$ level with  holes causes the low-energy PL line showing a long radiative recombination time. On the contrary, the high-energy PL edge is due to recombination of hot electrons with $E>E_p$, whose lifetime is limited by the energy relaxation of 20~ps. 

From these considerations, it follows that the sharpness of the high energy PL spectrum edge is determined by both the hole Fermi level smoothing ($<1$~meV at $T=2$~K) and the energy of the emitted phonons $\sim0.2$~meV. The width of the spectral distribution $n(E-E_p)$ below the percolation threshold for the electrons with $E<E_p$ is difficult to determine from our measurements, since it is masked by the wide distribution of equilibrium holes at the acceptors. It can be rather narrow since the electrons necessarily tunnel during relaxation down to stronger localized states. The approximate energy diagram is illustrated in Fig.~\ref{Scheme}(b), where the line denotes the smooth density of states and the hatched areas denote the occupied states.

There is one question that has been discussed in literature starting from the middle of the last century, namely concerning the location of the hole Fermi level in heavily $p-$doped GaAs: Is it inside the valence band or within the band gap? \cite{Pankove1965} Different opinions have been suggested, based on absorption and PL measurements in GaAs with varying impurity concentrations \cite{CardonaPRB1980, Zhang1995}. From our experiments it is difficult to conclude about the exact location of the hole Fermi level. This is, however, not critical for understanding the physics responsible for the step-like energy distribution of electrons studied here.

\section*{Conclusions}

To understand the electronic processes occurring near the Fermi-like tail of the PL spectrum of a strongly $p$-doped GaAs crystal a broad spectrum of optical techniques was applied. We found that the free electrons in the conduction band are separated from the continuum of the localized states by a particular percolation threshold. The presence of such a well defined electron level is a nontrivial fact due to the high disorder in the crystal lattice and the absence of translation symmetry. The percolation threshold results in the sharp step in the high energy photoluminescence tail. The spectral profile and its sharpness are determined not only by the hole distribution but also by the energy distribution of conduction band electrons, which exhibits a step-like behavior. It might be interesting to go to sub-Kelvin temperatures and disclose the physics affecting the sharpness of the PL spectral edge in more detail. Distinction between the contributions from the hole Fermi level and the electron percolation threshold requires further experimental efforts.

Another interesting property is that the electron spin lifetime is strongly limited by the short radiative recombination of the electrons themselves. This causes difficulties to measure the spin relaxation time, which is much longer and apparently exceeds nanoseconds.

\section*{Acknowledgements}

The authors thank A.S. Terekhov for providing the samples. I.A.A., D.R.Y., and M.B. acknowledge the Deutsche Forschungsgemeinschaft for financial support through the International Collaborative Research Centre TRR160 (Project No. A3). S.V.P. thanks the Russian Foundation for Basic Research for partial financial support (Research Grant No. 19-52-12046 NNIO\_a) and acknowledges St. Petersburg State University for the Research Grants Nos. 11.34.2.2012 and ID 40847559.


\end{document}